\documentclass[twocolumn,prl,aps,showpacs]{revtex4}%
\usepackage{amsfonts}
\usepackage{amssymb}
\usepackage{epsfig}
\usepackage{amsbsy}
\usepackage{amssymb}
\usepackage{amsmath}
\usepackage{graphicx}
\usepackage{graphics}
\usepackage{xcolor}
\usepackage[colorlinks=true,linkcolor=blue,citecolor=blue]{hyperref}%
\begin{document}
\title{Regularization of the spectral singularity in $\mathcal{PT}$-symmetric systems
by all order nonlinearities: nonreciprocity and optical isolation}
\author{Xuele Liu$^{1,2}$}
\altaffiliation{xuele@okstate.edu}

\author{Subhasish Dutta Gupta$^{1,3}$}
\author{G.S. Agarwal$^{1}$}
\affiliation{$^{1}$Department of Physics, Oklahoma State University, Stillwater, Oklahoma
74078, USA}
\affiliation{$^{2}$Department of Physics, University of Texas at Dallas, Richardson, TX
75080 USA}
\affiliation{$^{3}$School of Physics, University of Hyderabad, Hyderabad-500046, India}

\date{\today }

\begin{abstract}
Spectral singularities are ubiquitous with PT-symmetry leading to infinite
transmission and reflection coefficients. Such infinities imply the
divergence of the fields in the medium thereby breaking the very assumption of
the linearity of the medium used to obtain such singularities.
We identify saturable nonlinearity retaining contributions from all orders of
the field to limit the infinite growth and regularize the spectral
singularity. We present explicit numerical results to demonstrate
regularization. The all order nonlinear PT-symmetric device is shown to exhibit very
effective isolation or optical diode action, since
transmission through such a system is nonreciprocal. In contrast, a linear
system or a system with Kerr nonlinearity is known to have only reciprocal
transmission. Further we demonstrate optical bistability for such a system
with high contrast.

\end{abstract}

\maketitle


Complex $\mathcal{PT}$-symmetric potentials \cite{bender} possessing real
spectra have drawn considerable attention in recent times
\cite{ruter2010,v1,v2,v3,v4,ptreview}. Optics has played a very fertile role
to offer both theoretical and experimental testbeds for the unique properties
of $\mathcal{PT}$-symmetric systems
\cite{ruter2010,guo2009,longhi2010,periodic1}. The advantages and versatility
of optics stem from the one to one correspondence between the Schr\"{o}dingier
equation and the Helmholtz equation, and the ability of the dielectric
function to mimic the complex potential in certain cases. Though most of the
studies are devoted to laterally coupled systems (mostly waveguides and
directional couplers) \cite{ruter2010,guo2009,bloch}, there have been studies
on purely one dimensional systems with adjoining loss and gain sections
\cite{jphysa,mostafa1,mostafa3,ahmed1,invisibility}. Diverse effects like
power oscillations, loss-induced increase in transmission, non reciprocity in
reflection, invisibility etc have been reported
\cite{ruter2010,guo2009,periodic1}. It has been shown that the plasmonic
realization of $\mathcal{PT}$-symmetric devices hold a great deal of promise
for future technologies \cite{plasmonpt}. In view of the broad span of the
underlying physics concepts and their appeal for fundamental and applied
research a host of diverse systems has been studied to understand the
consequences of spontaneously broken $\mathcal{PT}$-symmetry. These include
Talbot imaging, coupled waveguides and periodic structures, Bloch
oscillations, chaos, coupled lumped networks, $\mathcal{PT}$-CPA lasers and
many other linear and nonlinear systems
\cite{pttalbot,periodic1,bloch,ptchaos,lrc1,ptcpa1,ptcpa2}.

Perhaps the most striking feature of the $\mathcal{PT}$-symmetric systems, is
their ability to possess spectral singularities which has been investigated in
detail \cite{mostafa1,mostafa2,ahmed1,mostafa3}. These studies assume that the underlying equations are linear. At the spectral singularities the
reflection and the transmission coefficients tend to infinity. There have been
nonlinear extensions of the theory involving a dispersive Kerr type
nonlinearity in a distributed feedback system \cite{nperiodic} reporting
bistability and non reciprocity in reflection. Soliton-type solutions and
other nonlinear modes have been reported in several other studies
\cite{soliton,nonlinearmodes}. Two very recent studies investigate the
robustness of the singularity in presence of Kerr type nonlinearity or in the
framework of a nonlinear Schr\"{o}dingier equation \cite{mostafa4,mostafa5}.
It is shown that the presence of a dispersive nonlinearity does not break the
parity-reflection symmetry of the spectral singularities, which are now power
dependent \cite{mostafa4}. In this letter we show that a different nonlinear mechanism is
needed to limit the infinite growth of the reflection and transmission
coefficients. In particular, with an example of a waveguide with equal finite
segments of gain and loss, we show that the incorporation of a saturation
mechanism for both gain and loss can lead to finite scattering amplitudes.
Indeed, a Kerr type nonlinearity can move around the location of the
singularity via an intensity-induced modification of the optical path. In view
of the divergence of the field for a linear system a nonlinearity having
contributions from all orders of the field is needed for the description of
the realistic system. Thus, a saturable nonlinearity can lead to an intricate
local field distribution affecting the intensity dependent real and imaginary
parts of the dielectric function, eventually breaking the $\mathcal{PT}%
$-symmetry. Note that saturable nonlinearities are extremely important in
optics and their role has been adequately discussed in the classic text by
Allen and Eberly \cite{eberly}. However, to the best of our knowledge,
saturation type nonlinearities have not been addressed in the context of
$\mathcal{PT}$-symmetric systems. Moving from dispersive to saturation type
nonlinearites increases the complexity of the problem to the extent that
analytical treatment is no longer possible and one has to revert to numerical
simulation. Note that exact or approximate solutions are known for dispersive
nonlinearity \cite{sdgreview}, while such are missing for saturable
active/passive media.

The proposed system has several definite advantages. Most importantly, such a device can act like a near-perfect isolator allowing only one way traffic \cite{isolator}. Similar optical diode
action can find many applications in chip-level optical circuitry. Indeed in the nonlinear regime it allows light to pass through only in
one direction implying non reciprocity in transmission. Note that the linear
counterpart can never have nonreciprocity in transmission, though reflection
can be nonreciprocal \cite{nonreciprocity,ahmed2}. Even in a Kerr nonlinear system
such nonreciprocity in transmission is absent \cite{nperiodic}. Secondly, as expected from a nonlinear system, there can be multivalued response with
high contrast between the `off' and `on' states. Needless to mention that
these feature are quite attractive for pure optical switching and logic
operations. 
.\begin{figure}[tb]
\includegraphics[width=0.45\textwidth]{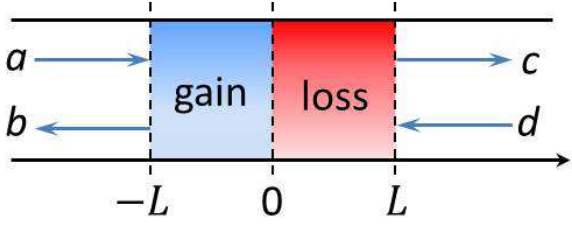}
\caption{Schematic view of the $\mathcal{PT}$-symmetric waveguide with equal segments of gain and loss medium.}%
\label{Fig1}%
\end{figure}

\textbf{$\mathcal{PT}$-symmetric system with all order nonlinearity}: Consider
the quasi-one dimensional system shown in Fig. \ref{Fig1}, consisting of
alternating equal segments of loss and gain. For a general power-dependent
nonlinear response the Helmholtz equation for such a system can be written as:%

\begin{equation}
\left[  \frac{\partial^{2}}{\partial x^{2}}+k^{2}\epsilon\left(  \left\vert
\Psi\right\vert ^{2}\right)  \right]  \Psi=0,\label{q1}%
\end{equation}
where $k=\omega/c$ is the vacumm wave vector and $\epsilon$ is the power
dependent dielectric function. We further model the loss (gain) medium to be a
collection of two-level atoms without (with) interaction. The dielectric
function can then be expressed as \cite{bk1}:%
\begin{equation}
\epsilon\left(  \left\vert \Psi\right\vert ^{2}\right)  =1+\chi_{0}\eta\left(
\left\vert \Psi\right\vert ^{2}\right)  ,\label{q2}%
\end{equation}
with%
\begin{equation}
\eta\left(  \left\vert \Psi\right\vert ^{2}\right)  =%
\begin{cases}
\frac{\delta+\text{i}}{1+\delta^{2}+\alpha\left\vert \Psi\right\vert ^{2}} &
0<x<L\\
\frac{\delta-\text{i}}{1+\delta^{2}+\alpha\left\vert \Psi\right\vert ^{2}} &
-L<x<0
\end{cases}
,\label{q3}%
\end{equation}
where $\delta=\left(  \omega-\omega_{0}\right)  /\tilde{\Gamma}$ is the
detunning normalized to the total decay rate $\tilde{\Gamma}=\gamma+\Gamma$
with $2\gamma$ and $\Gamma$ giving the Einstein $A$ coefficient and the
collisional line width, respectively. $\chi_{0}$ is the imaginary part of the
linear susceptibility at resonance. Since $\tilde{\Gamma}/\omega_{0}$ is
typically $\sim10^{-4}$, one has $\omega=\left(  1+10^{-4}\delta\right)
\omega_{0}$. In Eq. (\ref{q3}) we introduce a binary switch parameter $\alpha$
with values \textbf{1} and \textbf{0} to indicate the presence and absence of
nonlinearity. It is clear from a close inspection of Eq. (\ref{q3}), that an
increase in power affects both the real and imaginary parts of $\eta$. And,
thus, the propagation aspects determined by Eq. (\ref{q1}).

An indepth nonlinear study requires a complete understanding of the linear
properties of the system. Properties of the linear counterpart is now well
understood \cite{mostafa1,ptcpa1}. We recall some of the essential features
for self consistency. For illumination from both sides by plane waves with
amplitude $a$ and $d$, leading to scattered amplitudes $b$ and $c$, one can
relate them through a $2\times2$ matrix $\mathcal{M}$ as:%
\begin{equation}
\left[
\begin{array}
[c]{c}%
c\\
d
\end{array}
\right]  =\mathcal{M}\left[
\begin{array}
[c]{c}%
a\\
b
\end{array}
\right]  . \label{q4}%
\end{equation}
Particular cases of left-incident and right-incident scattering solutions can
be written as \cite{mostafa1}
\begin{equation}
\Psi^{l}\left(  x\right)  =%
\begin{cases}
c_{l}\left(  \text{e}^{\text{i}kx}+r_{l}\text{e}^{-\text{i}kx}\right)  &
x<-L\\
c_{l}t_{l}\text{e}^{\text{i}kx} & x>L
\end{cases}
, \label{q5}%
\end{equation}
and%
\begin{equation}
\Psi^{r}\left(  x\right)  =%
\begin{cases}
c_{r}t_{r}\text{e}^{-\text{i}kx} & x<-L\\
c_{r}\left(  \text{e}^{-\text{i}kx}+r_{r}\text{e}^{\text{i}kx}\right)  & x>L
\end{cases}
, \label{q6}%
\end{equation}
where $I_{l}=\left\vert c_{l}\right\vert ^{2}$ ($I_{r}=\left\vert
c_{r}\right\vert ^{2}$) are the incident light intensities, $r_{l}$, $t_{l}$
($r_{r}$, $t_{r}$) are the corresponding amplitude reflection coefficient and
amplitude transmission coefficient. We can then get transmission and
reflection coefficients $T_{l}=\left\vert t_{l}\right\vert ^{2}$
($T_{r}=\left\vert t_{r}\right\vert ^{2}$) and $R_{l}=\left\vert
r_{l}\right\vert ^{2}$ ($R_{r}=\left\vert r_{r}\right\vert ^{2}$).

In terms of the elements of $\mathcal{M}$, for the \textit{linear} system, the
amplitude transmission and reflection coefficients are given by matrix
elements of $\mathcal{M}$ \cite{mostafa1,ptcpa1}:%
\begin{equation}
t_{l}=\frac{\det\mathcal{M}}{\mathcal{M}_{22}}%
\begin{array}
[c]{cc}%
, &
\end{array}
t_{r}=\frac{1}{\mathcal{M}_{22}}%
\begin{array}
[c]{cc}%
, &
\end{array}
r_{l}=-\frac{\mathcal{M}_{21}}{\mathcal{M}_{22}}%
\begin{array}
[c]{cc}%
, &
\end{array}
r_{r}=\frac{\mathcal{M}_{12}}{\mathcal{M}_{22}}, \label{q7}%
\end{equation}
and we have $ \det\mathcal{M} = \mathcal{M}%
_{22}\mathcal{M}_{11}-\mathcal{M}_{12}\mathcal{M}_{21} \equiv1$
\cite{ar10}, which gives $t_{l}=t_{r}$ \cite{nonreciprocity,ahmed2}. For a
general proof directly based on Helmholtz equation, see equation 9 in
\cite{nonreciprocity}, which holds for any linear medium . For $\mathcal{PT}$
symmetric systems, one has additional relations $\mathcal{M}_{22}\left(
\omega\right)  =\mathcal{M}_{11}^{\ast}\left(  \omega^{\ast}\right)  $. Based
on these properties, for a real $\omega$, if $\left\vert t_{l}\right\vert
=\left\vert t_{r}\right\vert =1$, we have $\mathcal{M}_{22}\mathcal{M}%
_{11}=\left\vert \mathcal{M}_{22}\right\vert ^{2}=1$. In order to satisfy
$\det\mathcal{M}\left(  \omega\right)  \equiv1$, we need $\mathcal{M}%
_{12}\mathcal{M}_{21}=0$. Thus at least one of the amplitude reflection
coefficients $r_{l}$ and $r_{r}$ must be zero as will be confirmed by Fig. \ref{Fig2} below.
\begin{figure}[tb]
\includegraphics[width=0.45\textwidth]{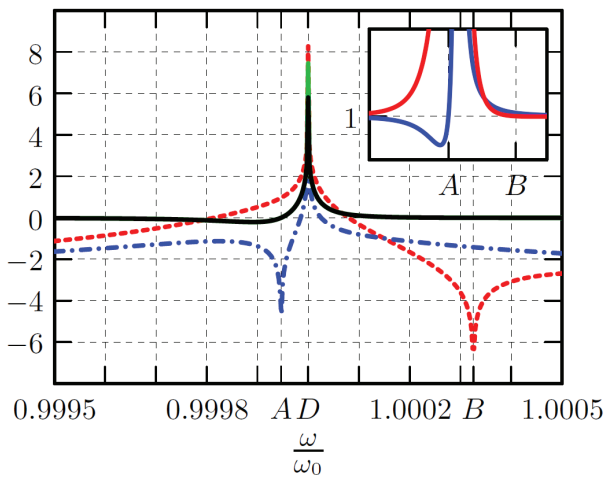}\caption{
Linear response of the $\mathcal{PT}$-symmetric system. 
 Black (darkgreen) solid line:
$\log_{10}{T_{l}}=\log_{10}{T_{r}}$; red dashed line: $\log_{10}{R_{l}}$; blue
dash-dotted line: $\log_{10}{R_{r}}$. Besides the singularity at $D$, the two critical points $A$ and $B$ occur at $\delta=-0.533$ and $\delta=3.256$, respectively. The inset shows the total scattered
intensity $S=R+T$ for left (blue solid line) and right (red dashed line) incidence.}%
\label{Fig2}%
\end{figure}
\par
For calculations we picked two singularities, namely, $n=0$ and $n=10$ of the
Table I in Ref. \cite{mostafa1}. Most of the results will refer to the most
discussed in literature case of $n=0$ as proof of principle, while the case of
$n=10$ will be discussed in the context of practical realization and viability
of a near perfect isolator in the optical domain. $n=0$ singularity occurs at
$\chi_{0}=1.82765566$ and from transition frequency $\omega_{0}$, we may get
the half-length of scattering area $L_{0}$ for divergence by the condition
$k_{0}L_{0}=1.06468255$, where $k_{0}=\frac{\omega_{0}}{c}$ is the wave vector
corresponding to the transition frequency. The results for linear response
(the various transmission and reflection coefficients) are shown in Fig.
\ref{Fig2}. The occurrence of the singularity is marked by letter $D$, while
the critical points are labeled by $A$, $B$. The one marked by $D$ reflects
the situation where all the reflection and transmission amplitudes are
singular, and it has been discussed in great detail \cite{mostafa1}. The
critical points in $R_{l}$ and $R_{r}$ occur at $A$ and $B$, respectively, for
left and right incidence. One can also demarcate parameters regimes of overall
gain and loss, looking at the total scattering $S=T+R$ as a function of
$\delta$. The gain (loss) domains is characterized by $S>1$ ($S<1$). For
example, for the critical point in $R_{l}$ at $A$, we notice that a crossover
at $\omega=\omega_{A}$ takes place from loss to gain behavior. Analogous
behavior is observed at the critical point $B$ for right incident waves. Note
that such cross-over behavior and both linear and nonlinear response in terms
of the criticality could be very different for a different resonance, since
they refer to different system sizes as compared to the wavelength. This will
be further highlighted in the context of the nonlinear response.
\par
\textbf{Regularization of spectral singularities}:
Having discussed the linear properties, we ask the question how nonlinearity
(see Eq. (\ref{q3})) affects the spectral singularities and critical points.
We concentrate mostly on the singularity at $D$, where all the coefficients
diverge in the linear system (see Fig. \ref{Fig2}.). In the context of the
zeroes of $R$ (the critical points), we show how the domain boundaries between
gain and loss get affected by the nonlinear response. We employ a variation of
the finite difference method for numerical integration of Eq. (\ref{q1}), with
nonlinearity (\ref{q3}) \cite{ap1}. For the power levels used in the
calculations, convergence was easily achieved.
\begin{figure}[t]
\includegraphics[width=0.45\textwidth]{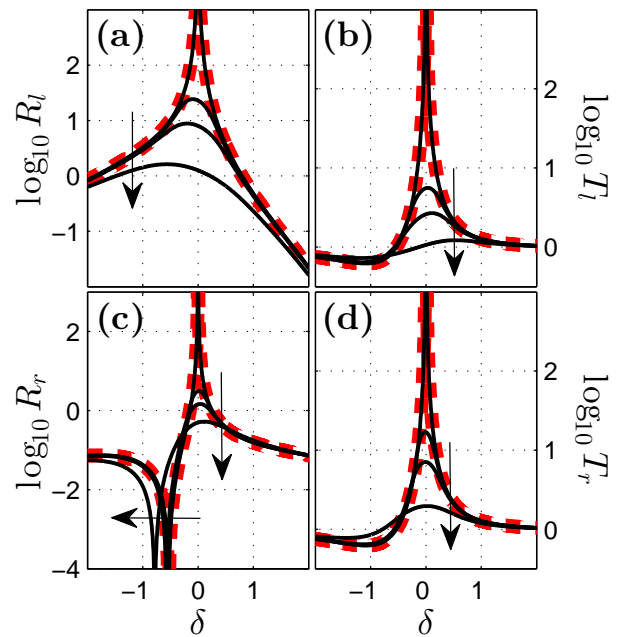}\caption{
Regularization of the spectral singularity. Intensity reflection ((a), (c)) and transmission coefficients ((b), (d)) as functions of normalized detuning $\delta$. Top (bottom) rows are for left and right incidence, respectively. Heavy red dashed line in each panel
gives the linear results ($\alpha=0$) for reference. The different curves are for increasing incident intensity $I=\left\vert
c\right\vert ^{2}$=$10^{-6}$, $0.01$, $0.04$ to $0.36$ along the direction of the arrows.}%
\label{Fig3}%
\end{figure}
\par
 Our central result concerns the ability of the saturation type
nonlinearity to limit the infinite growth, which eventually can lead to
optical isolation and diode action. The results for the reflection and
transmission coefficients for both left and right incidence for different
power levels are shown in Fig. \ref{Fig3}. Indeed, the singularity gets
destroyed by the saturation mechanism, as would be the case in a realistic
system. In fact, the peak amplitude gets less and less with increasing power
level. Moreover, there is a frequency shift of the peak for higher powers with
a broadening of the response.
 \begin{figure}[ttb]
\includegraphics[width=0.48\textwidth]{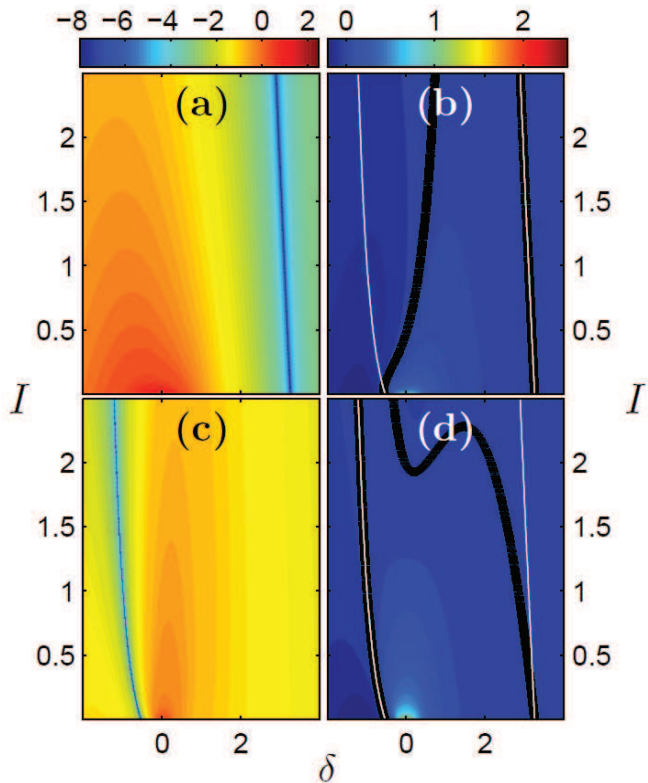}
\caption{
The color density images (in log scale) of the intensity reflection and transmission coefficients as functions of normalized detuning $\delta$ and incident intensity $I$: (a) $\log_{10}R_{l}$, (b) $\log
_{10}T_{l}$, (c) $\log_{10}R_{r}$ and (d)  $\log_{10}T_{r}$. The zeros of the
reflection coefficients $R_{r}$ and $R_{l}$ are shown by blue lines. For comparison, they are reproduced as thin lines in the right panels. The contour $T_{l}=1$ ($T_{r}=1$) is shown as heavy (black) lines in panel (b) (panel (d)).}
\label{Fig4}%
\end{figure}
\par
\textbf{Nonreciprocity, optical isolation and
bistability}: We now show how the gain/loss domain boundary given by $S=1$ gets distorted by
increasing power levels. The color density plot in Fig. \ref{Fig4} shows the
dependence of the transmission and reflection coefficients as functions of
$\delta$ and the incident intensity $I$. The domain
boundaries for left and right incidence characterized by $R_{l}=0$ and
$R_{r}=0$ are shown by blue lines in (a) and (c), respectively. The
case $T=1$ are depicted by the black heavy lines in (b) and (d) for left and
right incidence. It is clear that the effects of nonlinearity is not the same
for left and right incidence even in case of transmission. Recall that
transmission is known to be reciprocal in linear systems, and, as has been
shown recently, even in nonlinear systems with Kerr type nonlinearty
\cite{nonreciprocity,ahmed2,nperiodic}. In our case nonlinearty can lift the
parity degeneracy, and make transmission nonreciprocal. In fact, we show
below, that this nonreciprocity can be amplified to the extent that system
allows predominantly only one way transmission (optical isolation). It allows
light to pass through for left incidence while it is blocked for light coming
from the other side leading to the optical diode action. Another interesting
feature that need be noted from the black line in Fig. \ref{Fig4}(b) is that
the same power level corresponds to three distinct values of $\delta$ (say at
$I=2.2$). This is a typical signature of a nonlinear system and normally gets
manifested in bistable or multi-stable response. In what follow we show both
diode action and bistability in our system. To this goal we choose a system
with $L=3L_{0}$, which is detuned from the spectral singularity which occurs at
$k_{0}L_{0}=1.06468255$. We show that bistable response of transmission
coefficient for left incidence $T_{l}$ as a function of the modulus of the
incident amplitude. One can easily see the high contrast of the lower and
upper branches. The part of the curve with negative slope is unstable as in
standard hysteretic response. 
\begin{figure}[tbh]
\includegraphics[width=0.45\textwidth]{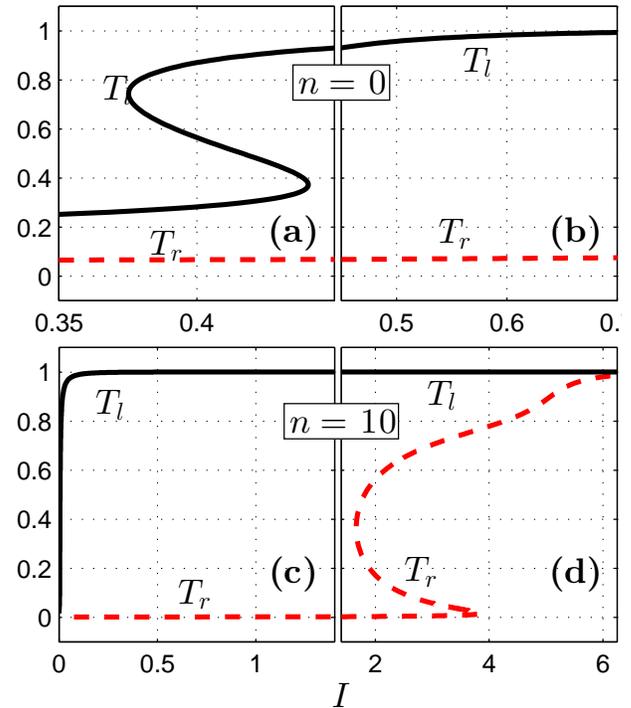} 
\caption{Optical isolation and bistability in
the nonlinear $\mathcal{PT}$ symmetric system when $L$ is no more the resonant
length $L_{0}$. (a-b) $L=3L_{0}$ for $n=0$, (c-d) $L=2.2L_{0}$ for $n=10$. The parameters for $n=0$ ($n=10$) were taken as $k_{0}L_{0}=1.06468255$, $\chi_{0}=1.82765566$ ($k_{0}L_{0}=32.8243878$, $\chi_{0}=0.17167639$)
as  in table I of \cite{mostafa1}.}%
\label{Fig5}%
\end{figure}
\par
The results discussed above clearly demonstrate the
remarkable potentials of $\mathcal{PT}$-symmetric systems with saturable nonlinearity.
However, in the optical domain $k_{0}L_{0}=1.06468255$ (for $n=0$) corresponds
to a system of size $100~nm-1~\mu m$, and it may be really challenging to create gain and loss regions. In order to overcome such difficulties, one can choose a larger system, say for $n=10$
with $k_{0}L_{0}=32.8243878$, $\chi_{0}=0.17167639$. With atomic vapors such a
set of parameters should be achievable. As mentioned earlier, the linear
and nonlinear response could be very different for different order
singularities with different $n$ values. The results for $n=10$ are shown in Figs. \ref{Fig5}(c) and \ref{Fig5}(d). A comparison of these with Figs. \ref{Fig5}(a) and \ref{Fig5}(b) for $n=0$ reveals
that one now has a completely contrasting result, since cases of left and right incidence exchange their roles. Now one has bistable response for $T_{r}$, while one had multi valued response for left incidence for $n=0$. Even for nominal intensities below a
certain threshold, one has near perfect optical isolation (Fig. \ref{Fig5}(c)). One also
has the familiar bistable response for larger power levels (Fig. \ref{Fig5}(d)). Clearly such effects open up new application possibilities of $\mathcal{PT}$-symmetric systems.
\par
In conclusion, we have studied a $\mathcal{PT}$-symmetric waveguide with
saturable nonlinearity. We show that the saturable nonliearity can
regularize the spectral singularities and limit the infinite growth of the
scattering amplitudes. Besides, the nonlinear response can exhibit
near-perfect optical isolation, showing near-total transmission for left incidence,
while it is close to zero for right incidence. We stress that nonreciprocity
in transmission is a property of our all order nonlinear $\mathcal{PT}$-symmetric system and is absent in linear transmission or even in transmission
through Kerr-nonlinear systems. We highlight the application potentials of
this optical diode action with near-perfect isolation and bistability in logic and memory devices.



\end{document}